\documentclass[aps,onecolumn,showpacs,footinbib]{revtex4}%
\usepackage[latin9]{inputenc}
\usepackage{amssymb}
\usepackage{amsmath}
\usepackage{amscd}
\usepackage{graphicx}
\usepackage{subfigure}
\usepackage{float}
\begin{document}
\draft
\title{Generation of atomic and field squeezing by adiabatic passage and symmetry
breaking}
\author{Shi-Biao Zheng}
\address{Department of Physics\\
Fuzhou University\\
Fuzhou 350002, P. R. China}
\date{\today }

\begin{abstract}
We propose an efficient scheme for realizing squeezing for both an atomic
ensemble and a cavity field via adiabatic evolution of the dark state of the
atom-cavity system. Controlled symmetry breaking of the Hamiltonian ensures
a unique dark state for the total system, in which the atomic system or
cavity mode is squeezed depending upon the choice of the detunings. Since
the generation of the atomic squeezed state requires neither the cavity mode
nor the atomic system to be excited, the decoherence effects are effectively
suppressed. The scheme is insensitive to the uncertainty in the atomic
number and imperfect timing, and the time needed for the generation of the
desired squeezed state decreases as the size of the system grows. The
required experimental techniques are within the scope of what can be
obtained in the present cavity QED setups.
\end{abstract}

\pacs{PACS number: 42.50.Dv, 03.67.Bg, 42.50.Pq}

\vskip 0.5cm \maketitle \narrowtext

\section{INTRODUCTION}

In recent years, there has been growing interest in quantum states from both
fundamental and practical points of view. Besides providing possibilities
for the test of fundamental quantum theory, nonclassical states have
potential applications. Of special interest are the squeezed states of an
electromagnetic field, whose quantum fluctuation in one quadrature is
reduced below the vacuum level at the expense of amplifying the noise in the
other quadrature [1]. Such states may be used to improve the signal-to-noise
ratio in optical communications [2] and detect gravitational wave [3]. In
correlated many-particle systems, spin squeezing can be defined as the
reduction of fluctuation in one collective spin component below the standard
quantum limit of the coherent spin state, at the expense of amplifying
another component. Such states are useful for atomic interferometers [4,5]
and high-precision spectroscopy [6,7]. Recently, it has been found that spin
squeezing is closely related to entanglement, which is the key resource for
quantum communication [8-10] and quantum computation [11]. Ulam-Orgikh and
Kitagawa have shown that spin squeezing implies pairwise entanglement [12],
and Wang and Sanders have given a quantitative relation between the
squeezing and concurrence for symmetric multispin states [13]. As the
pairwise entanglement is manifested only in the collective properties of the
multiqubit system, it is robust against the lose of coherence for a single
qubit, which is important for quantum information processing. Schemes have
been proposed for spin squeezing with optical lattice [14] and Bose-Einstein
condensates [15], and weak spin squeezing has been experimentally realized
[16,17].

Cavity QED is a qualified candidate for quantum state engineering and
quantum information processing. The strong atom-cavity coupling achievable
in a high-finesse cavity allows the generation of various nonclassical
states before decoherence sets in. The high degree of control over single
atoms [18-20] and atomic ensembles [21,22] in a resonator opens the
possibilities ranging from quantum state engineering and quantum networking
to quantum phase transitions. So far, schemes for preparing squeezed states
in cavity QED have been based on either parametric down conversion [23-25]
or quantum reservoir engineering [26-28]. To our knowledge, none of these
schemes has been experimentally realized. In this paper, we propose an
adiabatic passage scheme for generation of squeezed states for both the
atomic system and cavity mode. Unlike previous schemes, our scheme is based
on the symmetry breaking of the Hamiltonian for the combined atom-cavity
system, which ensures a unique dark state of the interaction Hamiltonian,
given by the product of the squeezed state of the atomic system with the
vacuum state of the cavity mode, or vice versa. The squeezing parameter is
controllable via the intensities and phases of the classical driving fields.
Compared with the previous schemes, the scheme has the following important
features: (i) For the generation of the atomic squeezed state, neither the
cavity mode nor the atomic system is excited so that the model is robust
against decoherence mechanisms and a high-fidelity squeezed state can be
generated beyond the strong coupling regime; (ii) the method is immune to
the uncertainty in the atomic number; (iii) the interaction time need not to
be accurately adjusted as long as the adiabatic condition is fulfilled; and
(iv) the time needed to produce the state with a desired squeezing parameter
decreases when the number of atoms increases. The scheme is feasible with
current experimental technology.

\section{EFFECTIVE HAMILTONIAN}

We consider that $N$ atoms are trapped in a single-mode cavity. The atomic
level configuration is shown in Fig. 1. Each atom has two excited states $%
\left| r\right\rangle $ and $\left| s\right\rangle $ and two ground states $%
\left| e\right\rangle $ and $\left| g\right\rangle $. The cavity mode
couples to the transitions $\left| g\right\rangle \longleftrightarrow \left|
r\right\rangle $ and $\left| e\right\rangle \longleftrightarrow \left|
s\right\rangle $ with the coupling strengths $g_1$ and $g_2$, respectively.
Meanwhile, one laser couples to the transition $\left| e\right\rangle
\longleftrightarrow \left| r\right\rangle $ with Rabi frequency $\Omega _1$
and phase $\phi _1$ and another laser couples to $\left| g\right\rangle
\longleftrightarrow \left| s\right\rangle $ with Rabi frequency $\Omega _2$
and phase $\phi _2$. As will be shown, these fields are used to drive two
distinct Raman transitions between two atomic ground states, which lead to
the competition between the annihilation and creation operators of the
collective atomic mode or of the cavity mode, making the atomic or field
squeezed state be the unique dark state of the effective two-mode coupling
Hamiltonian depending upon the choice of the field detunings. The
Hamiltonian for the system is ($\hbar =1$)
\begin{eqnarray}
H &=&\omega _aa^{\dagger }a+\sum_{j=1}^N(g_1a\left| r_j\right\rangle
\left\langle g_j\right| +\Omega _1e^{i\phi _1}e^{-i\omega _1t}\left|
r_j\right\rangle \left\langle e_j\right|  \nonumber \\
&&+g_2a\left| s_j\right\rangle \left\langle e_j\right| +\Omega _2e^{i\phi
_2}e^{-i\omega _2t}\left| s_j\right\rangle \left\langle g_j\right| +H.c. \\
&&+\omega _s\left| s_j\right\rangle \left\langle s_j\right| +\omega _r\left|
r_j\right\rangle \left\langle r_j\right| +\omega _e\left| e_j\right\rangle
\left\langle e_j\right| ),  \nonumber
\end{eqnarray}
where $\omega _s$, $\omega _r$, and $\omega _e$ are the energies of the
levels $\left| s\right\rangle $, $\left| r\right\rangle $, and $\left|
e\right\rangle $, $\omega _a$ is the frequency of the cavity mode, and $%
\omega _1$ and $\omega _2$ are the frequencies of the two classical fields.
Here the energy of level $\left| g\right\rangle $ is set to zero. We now
switch to the interaction picture with respect to
\begin{equation}
H_0=\omega a^{\dagger }a+\sum_{j=1}^N(\omega _2\left| s_j\right\rangle
\left\langle s_j\right| +\omega \left| r_j\right\rangle \left\langle
r_j\right| +\omega _e^{^{\prime }}\left| e_j\right\rangle \left\langle
e_j\right| ),
\end{equation}
where $\omega =(\omega _1+\omega _2)/2$ and $\omega _e^{^{\prime }}=(\omega
_2-\omega _1)/2$ are close to $\omega _c$ and $\omega _e$, respectively.
Then the Hamiltonian describing the atom-field interaction is

\begin{equation}
H_i=H_{i,0}+H_{i,1}+H_{i,2},
\end{equation}
where
\begin{eqnarray}
H_{i,0} &=&(\omega _a-\omega )a^{\dagger }a+(\omega _e-\omega _e^{^{\prime
}})\sum_{j=1}^N\left| e_j\right\rangle \left\langle e_j\right|  \nonumber \\
H_{i,1} &=&A_{i,1}+A_{i,1}^{\dagger }+\Delta _1\sum_{j=1}^N\left|
r_j\right\rangle \left\langle r_j\right| ,  \nonumber \\
H_{i,2} &=&A_{i,2}+A_{i,2}^{\dagger }+\Delta _2\sum_{j=1}^N\left|
s_j\right\rangle \left\langle s_j\right| ,  \nonumber \\
A_{i,1} &=&\sum_{j=1}^N(g_1a\left| r_j\right\rangle \left\langle g_j\right|
+\Omega _1e^{i\phi _1}\left| r_j\right\rangle \left\langle e_j\right| ),
\nonumber \\
A_{i,2} &=&\sum_{j=1}^N(g_2a\left| s_j\right\rangle \left\langle e_j\right|
+\Omega _2e^{i\phi _2}\left| s_j\right\rangle \left\langle g_j\right| ),
\end{eqnarray}
and $\Delta _1=\omega _r-\omega $ and $\Delta _2=\omega _s-\omega _2$ are
the detunings between the fields and the respective atomic transitions.

Under the large detuning condition, i.e., $\Delta _1$, $\Delta _2\gg g_1$, $%
g_2$, $\Omega _1$, $\Omega _2$, $\omega _a-\omega $, $\omega _e-\omega
_e^{^{\prime }}$, $H_{i,1}$ and $H_{i,2}$ can be respectively replaced by
the effective Hamiltonians [29]
\begin{eqnarray}
H_{{\sl eff,1}} &=&\Delta _1\sum_{j=1}^N\left| r_j\right\rangle \left\langle
r_j\right| +\frac{[A_{i,1},\text{ }A_{i,1}^{\dagger }]}{\Delta _1}  \nonumber
\\
&=&(\Delta _1+\eta _e+\xi _ga^{\dagger }a)\sum_{j=1}^N\left|
r_j\right\rangle \left\langle r_j\right| +\sum_{j,k=1}^N\xi _g\left|
r_j\right\rangle \left\langle g_j\right| \otimes \left| g_k\right\rangle
\left\langle r_k\right|  \nonumber \\
&&-\eta _e(S_z+N/2)-\xi _ga^{\dagger }a(N/2-S_z)-(\lambda _1e^{-i\phi
_1}aS^{+}+H.c.),
\end{eqnarray}
and
\begin{eqnarray}
H_{{\sl eff,2}} &=&\Delta _2\sum_{j=1}^N\left| s_j\right\rangle \left\langle
s_j\right| +\frac{[A_{i,2},\text{ }A_{i,2}^{\dagger }]}{\Delta _2}  \nonumber
\\
&=&(\Delta _2+\eta _g+\xi _ea^{\dagger }a)\sum_{j=1}^N\left|
s_j\right\rangle \left\langle s_j\right| +\sum_{j,k=1}^N\xi _e\left|
s_j\right\rangle \left\langle e_j\right| \otimes \left| e_k\right\rangle
\left\langle s_k\right|  \nonumber \\
&&-\xi _ea^{\dagger }a(S_z+N/2)-\eta _g(N/2-S_z)-(\lambda _2e^{i\phi
_2}a^{\dagger }S^{+}+H.c.),
\end{eqnarray}
where $S^{+}=\sum_{j=1}^N\left| e_j\right\rangle \left\langle g_j\right| $, $%
S_z=\frac 12\sum_{j=1}^N(\left| e_j\right\rangle \left\langle e_j\right|
-\left| g_j\right\rangle \left\langle g_j\right| )$, $\eta _e=\frac{\Omega
_1^2}{\Delta _1}$, $\eta _g=\frac{\Omega _2^2}{\Delta _2}$, $\xi _e=\frac{%
g_2^2}{\Delta _2}$, $\xi _g=\frac{g_1^2}{\Delta _1}$, $\lambda _1=\frac{%
\Omega _1g_1}{\Delta _1}$, and $\lambda _2=\frac{\Omega _2g_2}{\Delta _2}$.
The atomic excitation number are not changed during the interaction since
the atomic excitation number operator $\sum_{j=1}^N(\left| s_j\right\rangle
\left\langle s_j\right| +\left| r_j\right\rangle \left\langle r_j\right| )$
commutes with the total effective Hamiltonian $H_{i{\sl ,0}}+H_{{\sl eff,1}%
}+H_{{\sl eff,2}}$. When all the atoms are initially in the ground states,
they would remain in the ground states, i.e., the excited states $\left|
r\right\rangle $ and $\left| s\right\rangle $ can be adiabatically
eliminated. Since none of the operators $\left| r_j\right\rangle
\left\langle r_j\right| $, $\left| s_j\right\rangle \left\langle s_j\right| $%
, $\left| g_k\right\rangle \left\langle r_k\right| $, and $\left|
e_k\right\rangle \left\langle s_k\right| $ has any effect on the atomic
ground states, the terms containing each of these operators can be discarded
and the effective Hamiltonians $H_{{\sl eff,1}}$ and $H_{{\sl eff,2}}$
reduces to
\begin{equation}
H_{{\sl eff,1}}=-\eta _e(S_z+N/2)-\xi _ga^{\dagger }a(N/2-S_z)-(\lambda
_1e^{-i\phi _1}aS^{+}+H.c.),
\end{equation}
and
\begin{equation}
H_{eff,2}=-\xi _ea^{\dagger }a(S_z+N/2)-\eta _g(N/2-S_z)-(\lambda _2e^{i\phi
_2}a^{\dagger }S^{+}+H.c.).
\end{equation}
$H_{{\sl eff,1}}$ and $H_{{\sl eff,2}}$ describe two distinct Raman
transitions between the two atomic ground states.

\section{GENERATION OF SQUEEZED STATES}

In the Holstein-Primakoff representation, the collective spin operators \{$%
S_z,S^{\pm }\}$ are associated with the bosonic annihilation and creation
operators $b$ and $b^{\dagger }$ via

\begin{eqnarray}
S^{+} &=&b^{\dagger }\sqrt{N-b^{\dagger }b},S^{-}=\sqrt{N-b^{\dagger }b}b, \\
S_z &=&b^{\dagger }b-N/2.  \nonumber
\end{eqnarray}
When the average number of atoms in the state $\left| e\right\rangle $ is
much smaller than total atomic number, i.e., $\left\langle b^{\dagger
}b\right\rangle \ll N$, the collective spin operators are well approximated
by $S^{+}\simeq \sqrt{N}b^{\dagger }$, $S^{-}\simeq \sqrt{N}b$, and $%
S_z\simeq N/2.$ In this case the atomic ensemble can be regarded as a
bosonic system, and the transition of one atom from $\left| g\right\rangle $
to $\left| e\right\rangle $ corresponds to the creation of one quantum in
the effective bosonic mode, and vice versa. Then the effective Hamiltonians $%
H_{eff,1}$ and $H_{{\sl eff,2}}$ approximate to
\begin{equation}
H_{{\sl eff,1}}=-N\xi _ga^{\dagger }a-(\sqrt{N}\lambda _1e^{-i\phi
_1}ab^{\dagger }+H.c.),
\end{equation}
and
\begin{equation}
H_{{\sl eff,2}}=-N\eta _g-(\sqrt{N}\lambda _2e^{i\phi _2}a^{\dagger
}b^{\dagger }+H.c.).
\end{equation}
The dynamics of the system is given by the total effective Hamiltonian
\begin{eqnarray}
H_{eff} &=&H_{i,0}+H_{{\sl eff,1}}+H_{{\sl eff,2}}  \nonumber \\
&=&\delta _aa^{\dagger }a+\delta _bb^{\dagger }b-[\sqrt{N}(\lambda
_1ae^{-i\phi _1}+\lambda _2a^{\dagger }e^{i\phi _2})b^{\dagger }+H.c.],
\end{eqnarray}
where $\delta _a=\omega _a-\omega -N\xi _g$, $\delta _b=\omega _e-\omega
_e^{^{\prime }}$. We here have discarded the constant term. Set $\delta _b=0$
and $\delta _a\neq 0$. In this case the effective Hamiltonian reduces to
\begin{equation}
H_{{\sl eff}}=\delta _aa^{\dagger }a-[\sqrt{N}a^{\dagger }(\lambda
_1e^{i\phi _1}b+\lambda _2e^{i\phi _2}b^{\dagger })+H.c.].
\end{equation}
Perform the unitary transformation $\stackrel{\sim }{H}_{eff}=S_b^{\dagger
}(\xi )H_{eff}S_b(\xi )$ with the atomic squeezing operator $S_b(\xi
)=e^{(\xi ^{*}b^2-\xi b^{\dagger 2})/2}$, where $\xi =re^{i\theta }$. If we
choose the squeezing strength $r=\tanh ^{-1}\frac{\lambda _2}{\lambda _1}$
and squeezing phase $\theta =-(\phi _1+\phi _2)$, the transformed
Hamiltonian is given by
\begin{equation}
\stackrel{\sim }{H}_{eff}=\delta _aa^{\dagger }a-\mu (e^{-i\phi
_1}ab^{\dagger }+H.c.),
\end{equation}
where $\mu =\sqrt{N(\lambda _1^2-\lambda _2^2)}$. This Hamiltonian describes
the linear coupling between the field mode and the transformed collective
atomic mode, with the total quantum number being conserved. The dark state
(eigenstate with zero eigenenergy) of $\stackrel{\sim }{H}_{eff}$ is the
vacuum state $\left| 0\right\rangle _a\left| 0\right\rangle _b$. This
implies that the dark state of the effective Hamiltonian $H_{{\sl eff}}$ is $%
S_b(\xi )\left| 0\right\rangle _a\left| 0\right\rangle _b$, which is the
product of the squeezed atomic state with the vacuum field state. The
squeezing strength and squeezing phase are controllable by the Rabi
frequency $\Omega _2$ and phase $\phi _2$ of the second classical field.
Suppose that the atom-cavity system is initially in the vacuum state $\left|
0\right\rangle _a\left| 0\right\rangle _b$ and the Rabi frequency $\Omega _2$
is initially zero so that the initial state is identical to the dark state.
When the Rabi frequency $\Omega _2$ is slowly increased with the change rate
much smaller than the energy scales, adiabatic theorem ensure that the
system approximately follows the dark state, leading to the atomic squeezed
state $S_b(\xi )\left| 0\right\rangle _b$. During the adiabatic evolution,
neither the atomic system nor the cavity mode is excited and thus the method
is robust against decoherence. Furthermore, the squeezing parameter is
decided by the ratio between the two Raman coupling strengths and
independent of the number of atoms. This implies that the uncertainty in the
atomic number does not affect the produced state.

Now we proceeds to show how the squeezed state of the cavity mode can be
generated. Setting $\delta _a=0$ and $\delta _b\neq 0$ we obtain the
effective Hamiltonian
\begin{equation}
H_{{\sl eff}}^{^{\prime }}=\delta _bb^{\dagger }b-[\sqrt{N}(\lambda
_1ae^{-i\phi _1}+\lambda _2a^{\dagger }e^{i\phi _2})b^{\dagger }+H.c.].
\nonumber
\end{equation}
In this case we transform the effective Hamiltonian as $\stackrel{\sim }{H}_{%
{\sl eff}}^{^{\prime }}=S_a^{\dagger }(\zeta )H_{eff}^{^{\prime }}S_a(\zeta
) $ with the field squeezing operator $S_a(\zeta )=e^{(\zeta ^{*}a^2-\zeta
a^{\dagger 2})/2}$ , where $\zeta =re^{i(\phi _1+\phi _2)}$. We obtain the
new engineered effective Hamiltonian
\begin{equation}
\stackrel{\sim }{H}_{{\sl eff}}^{^{\prime }}=\delta _bb^{\dagger }b-\mu
(e^{-i\phi _1}ab^{\dagger }+H.c.).  \nonumber
\end{equation}
This Hamiltonian describes the linear coupling between the transformed field
mode and the collective atomic mode. The squeezed state of the cavity mode
can be produced from the vacuum state by adiabatically increasing the Rabi
frequency $\Omega _2$ from zero. It should be noted that if $\delta
_a=\delta _b=0$ both the states $S_a(\zeta )\left| 0\right\rangle _a\left|
0\right\rangle _b$ and $S_b(\xi )\left| 0\right\rangle _a\left|
0\right\rangle _b$ are the eigenstates with null eigenenergy of the
effective Hamiltonian. To ensure the required adiabatic change, it is
necessary to lift the degeneracy. A nonzero detuning $\delta _a$ or $\delta
_b$ breaks the symmetry of the effective Hamiltonian and renders $S_b(\xi
)\left| 0\right\rangle _a\left| 0\right\rangle _b$ or $S_a(\zeta )\left|
0\right\rangle _a\left| 0\right\rangle _b$ the unique dark state.

\section{DISCUSSION AND CONCLUSION}

We now address the experimental issues. We consider an ensemble of $N\sim
10^6$ $^{87}$Rb atoms trapped in a a ring cavity. The cavity mode is
linearly polarized along an axis perpendicular to an applied magnetic field.
The states $\left| e\right\rangle $ and $\left| g\right\rangle $ can be the
Zemman sublevels $\left| F=1,M_F=\pm 1\right\rangle $ of the ground state $%
\left| 5^2S_{1/2}\right\rangle $. Take $g_1\simeq g_2\simeq 2\pi \times 50$
kHz, $\kappa =2\pi \times 25$ kHz, and $\gamma =2\pi \times 6$ MHz [30,31],
where $\kappa $ and $\gamma $ are the cavity decay rate and the atomic
spontaneous emission rate, respectively. For $\Omega _1/\Delta _1=1/200$ we
have $\lambda _1=2\pi \times 0.25$ kHz. Slow variation of $\Omega _2/\Delta
_2$ from 0 to $1/250$ leads to the squeezing strength $r\simeq 1.1$. During
the adiabatic evolution the effective linear coupling $\mu $ between the two
bosonic modes is varied from $\mu _{\max }=2\pi \times 250$ kHz to $\mu
_{\min }=2\pi \times 150$ kHz. For the generation of the atomic squeezed
state the adiabatic approximation requires that $\delta _aT\gg 1$ and $%
\delta ET\gg 1$, where $\delta E=\sqrt{\mu ^2+\delta _a^2/4}-\delta _a/2$ is
the energy gap between the dark state and the nearest states of the
Hamiltonian $\stackrel{\sim }{H}_{eff}$ with nonzero eigenenergy. We note
that $\delta E$ decreases as $\delta _a$ increases. To satisfy the adiabatic
condition the value of $\delta _a$ should be moderate in comparison with
that of $\mu $. Choosing $\delta _a=\sqrt{\stackrel{-}{\mu }^2+\delta _a^2/4}%
-\delta _a/2$, where $\stackrel{-}{\mu }=(\mu _{\max }+\mu _{\min })/2$,
leads to $\delta _a=\sqrt{\stackrel{-}{\mu }/2}=2\pi \times 100$ kHz. Then
the energy gap $\delta E$ is varied from $\delta E_{\max }=2\pi \times 205$
kHz to $\delta E_{\min }=2\pi \times 108$ kHz. If we set $T=10/g\simeq 31.8$
$\mu $s, then the leakage error to the bright eigenstates is on the order of
$P_b\sim 1/(\delta E_{\min }T)^2+1/(\delta _aT)^2\simeq 4.64\times 10^{-3}$,
where $\delta \stackrel{-}{E}$ is the average value of $\delta E$. This
leads to the effective decoherence rate $\kappa _e\sim P_b\kappa \simeq 2\pi
\times 0.116$ kHz of the cavity mode. The atoms are virtually excited during
the evolution, the effective decoherence rate due to the atomic spontaneous
emission is $\gamma _e\sim \gamma \Omega _{2,\max }^2/(2\Delta _2^2)\simeq
2\pi \times 4.8\times 10^{-2}$ kHz. The error induced by decoherence is
about $(\kappa _e+\gamma _e)T\simeq 3.28\times 10^{-2}$. Therefore, both the
adiabaticity and neglect of decoherence can be perfectly satisfied. The
result shows that the atomic squeezed state with a high fidelity can be
generated even when the cooperativity parameter $g^2/2\gamma \kappa $ is as
low as $10^{-2}$. It should be noted that the interaction time does not need
to be adjusted very accurately as long as the adiabatic condition is
satisfied. For adiabatic following of the squeezed field state the cavity
quality needs to be improved. In recent experiments with a Bose-Einstein
condensate strongly coupled to an optical cavity [32,33], much higher
cooperativity parameters have been achieved, indicating a high-fidelity
squeezed state for the cavity can also be produced. An alternative way to
generate the squeezed field state is to use the matter-light state-transfer
scheme. After the atomic squeezed state has been produced, the resonant
Raman transition induced by the cavity mode and the laser field $\Omega _1$
can return all of the atoms to the state $\left| g\right\rangle $ and
transfer the state of the collective atomic mode to the cavity mode.

In conclusion, we have proposed a scheme for deterministically producing
squeezed states for both the collective atomic mode and the cavity mode via
adiabatic following of the dark state of the atom-cavity system by breaking
the symmetry of the Hamiltonian, which renders the dark state unique.
Compared with the previous methods, the present method offers potential
practical advantages. For the generation of atomic squeezed states, neither
the cavity mode nor the atomic system is excited, and the model is robust
against decoherence mechanisms. The state evolution is unaffected by the
uncertainty in the atomic number and imperfect timing. The scheme is within
reach of current experiments and expands the range of possibilities for
quantum state preparation in continuous variable systems.

This work was supported by the Major State Basic Research Development
Program of China under Grant No. 2012CB921601, the National Natural Science
Foundation of China under Grant No. 10974028, the Doctoral Foundation of the
Ministry of Education of China under Grant No. 20093514110009, and the
Natural Science Foundation of Fujian Province under Grant No. 2009J06002.

Fig.1 (color online) The atomic level configuration and excitation scheme.
Each atom has two ground states $\left| e\right\rangle $ and $\left|
g\right\rangle $ and two excited states $\left| r\right\rangle $ and $\left|
s\right\rangle $. The transitions $\left| g\right\rangle \longleftrightarrow
\left| r\right\rangle $ and $\left| e\right\rangle \longleftrightarrow
\left| s\right\rangle $ are coupled to the cavity mode with the coupling
strengths $g_1$ and $g_2$, respectively. Furthermore, the transitions $%
\left| e\right\rangle \longleftrightarrow \left| r\right\rangle $ and $%
\left| g\right\rangle \longleftrightarrow \left| s\right\rangle $ are driven
by two classical fields with the Rabi frequencies $\Omega _1$ and $\Omega _2$%
, respectively. The cavity mode, together with the laser fields, induce two
Raman transitions between the two atomic ground states.
\begin{figure}[C]
\includegraphics[width=0.5\columnwidth]{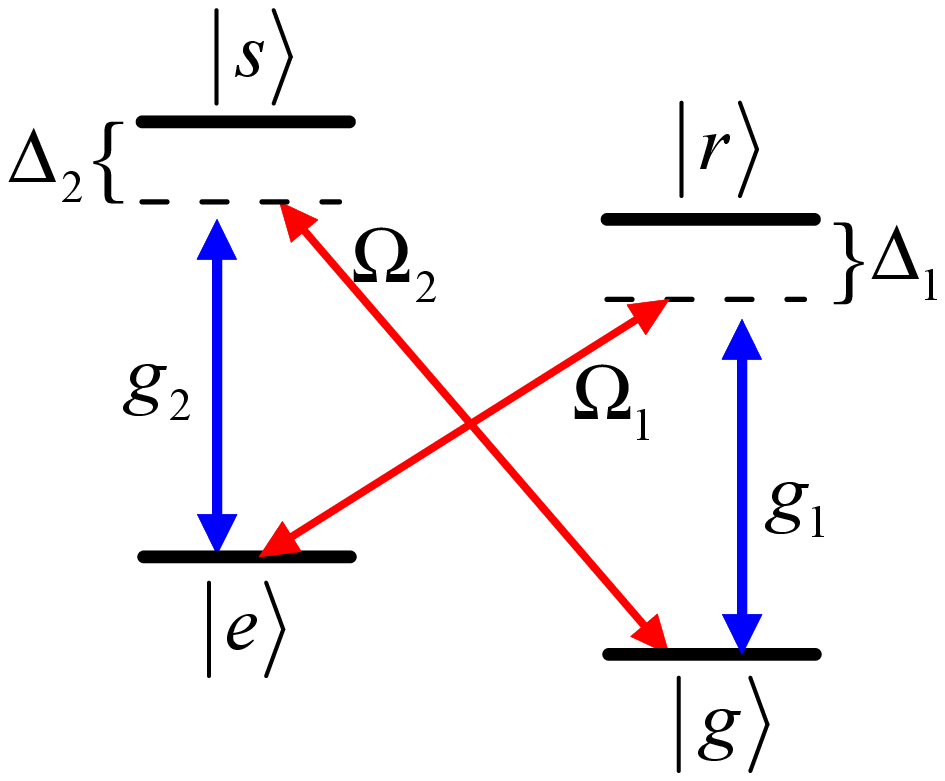}
\caption{}
\end{figure}
\end{document}